\documentclass[12pt]{article}
\usepackage{latexsym}
\usepackage{amssymb}
\usepackage{amsmath}
\usepackage{graphicx}
\newcommand{\tr}{\hbox{Tr}}
\newcommand{\gz}{g_{(0)}}
\newcommand{\gt}{g_{(2)}}
\newcommand{\gf}{g_{(4)}}
\newcommand{\gs}{g_{(6)}}
\newcommand{\ginv}{g^{-1}_{(0)}}

\begin{document}

\title{Stress-Energy Tensors for Higher Dimensional Gravity}
\author{{\small Andrew DeBenedictis
\footnote{e-mail: debened@death.phys.sfu.ca}}
\\ {\small K. S. Viswanathan \footnote{e-mail: kviswana@sfu.ca}}
\\ \it{\small Department of Physics}
\\ \it{\small Simon Fraser University, Burnaby, British Columbia, Canada V5A
1S6}}
\date{October 6, 1999}
\maketitle

\begin{abstract} We calculate, in the context of higher dimensional
gravity, the stress-energy tensor and Weyl anomaly associated with anti-de
Sitter and anti-de Sitter black hole solutions. The boundary counter-term
method is used to regularize the action and the resulting stress-energy
tensor yields both the correct black hole energies as well as a vacuum 
energy contribution which is interpreted as a Casimir energy. This 
calculation is done up to $d = 8$ ($d$ being the boundary
dimension). We confirm some results for $d < 8$ as well as comment on some
new results. All results for $d=8$ are new.

\end{abstract}

\vspace{3mm}
PACS number(s): 04.20.H, 04.70.-s, 04.62.+v \\

\section{Introduction}
\qquad There has been much debate in general relativity as to how to
assign the stress-energy contribution due to the gravitational field.
Early works in this field include Einstein's introduction of the pseudo-
tensor \cite{ref:estn} however, this lacks invariance which should be
present in a covariant theory such as relativity. Levi-Civita's
argument \cite{ref:levcit} that the stress-energy tensor, as defined
by the Einstein tensor, plays the role of ``balancing out'' space-time's
stress energy is a more natural interpretation. More recent work on 
the subject may be found in \cite{ref:babak}.

\qquad The motivation for the counterterm subtraction method as found in
\cite{ref:balakraus}, \cite{ref:krauslarsen} and \cite{ref:meyers} is
not so much to define a stress-energy tensor for gravity but to
regularize the gravitational action for spacetimes with constant energy
densities due to a cosmological term:
\begin{equation}
S=\frac{1}{16\pi}\int_{B}d^{D}x\,\sqrt{|g|}\left(R+
\alpha^{2}\left(d(d-1)\right)\right) -
\frac{1}{8\pi}\int_{\partial{B}}d^{d}x\sqrt{|\gamma|}K
+\frac{1}{8\pi}S_{(ct)}(\gamma). \label{eq:stdactn}
\end{equation}
\footnote{Conventions follow that of \cite{ref:MTW} and $\alpha$ is $1/l$
where $l$ is the radius of the anti-de Sitter spacetime.} Here, $d$ is the
dimension of the boundary ($\partial B$), of the D-dimensional bulk spacetime
($B$). $\gamma$ is the determinant of the boundary metric, $\gamma_{ab}$ and
$K$ is the trace of the extrinsic curvature, $K_{ab}$ of the boundary. The
first term is the usual Einstein Hilbert term, the second the Gibbons-Hawking
surface term and the third is the counterterm action which removes the
stress-energy tensor divergences which result from the previous terms.

\qquad Varying the first two terms with respect to the boundary metric
yields the unrenormalized stress-energy tensor \cite{ref:brnandyork}:
\begin{equation}
T^{a b}_{(unren)}=\frac{1}{8\pi}\left(K^{ab}-K\gamma^{ab}\right).
\label{eq:tunren}
\end{equation}

\qquad The final term in (\ref{eq:stdactn}) may be constructed in two ways. 
There is the
background subtraction method of Brown and York \cite{ref:brnandyork}
where one chooses for $S_{(ct)}$ the action of a spacetime with the same
intrinsic geometry as the spacetime of interest. For black holes of mass
$M$ for example, a natural choice would be the $M=0$ limit of the original
spacetime (for example see \cite{ref:horoandmeyers}). Another method,
which we will use here, involves constructing $S_{(ct)}$ from curvature
invariants of $\gamma^{ab}$ \cite{ref:balakraus} and therefore bulk
equations of motion will not be affected. This method allows definitions
of conserved quantities without the introduction of a spacetime which is
external to the one under study. Also, this method is useful when
considering  spacetimes which do not have a natural reference background
to which a comparison may be made or when non-trivial topologies are present.

\subsection{The boundary counterterm method.}
\qquad In this section we will briefly review the method of sucessive
boundary counterterms which was first introduced in
\cite{ref:balakraus} and \cite{ref:meyers}. Schematically, the
counterterms may be written as an expansion in inverse powers of $\alpha$:
\begin{equation}
S_{(ct)}=\alpha\left(S^{(0)}+\alpha^{-2}S^{(1)}+...\right).
\end{equation}
The resulting total stress-energy tensor is then just
\begin{equation}
T^{ab}_{(finite)}=T^{ab}_{(unren)}+T^{ab}_{(ct)},
\end{equation}
where $T^{ab}_{(ct)}$  comes from the variation of $S_{(ct)}$ with respect to
the boundary metric $\gamma_{ab}$.

The appropriate counterterms are uniquely determined by demanding that
the resulting stress-energy tensor be finite. This finite tensor must
then reproduce the correct conserved quantities for known solutions.

\qquad If we write the line element of the boundary in ADM form where the
hypersurfaces ($\Sigma$) are spacelike surfaces of constant $t$:
\begin{equation}
ds^2_{\partial
B}=-N^{2}_{\Sigma}dt^{2}+\sigma_{ab}(dx^{a}+N_{\Sigma}^{a}dt)(dx^{b}+N_{\Sigma}^{b}dt),
\label{eq:admmetric}
\end{equation}
then the energy of the spacetime is obtained from the energy density as
\begin{equation}
M=\int_{\Sigma}d^{p}x\sqrt{\sigma}\sqrt{|g_{tt}|}u^{a}u^{b}T_{ab(finite)}.
\label{eq:energy}
\end{equation}
Where $u^{a}$ is the unit normal to $\Sigma$.

\qquad Although we will be primarily interested in energies, other conserved
quantities may be similarly calculated by exploiting other Killing 
symmetries.
The integral in (\ref{eq:energy}) will diverge due to the asymptotic behaviour
of the metric determinant unless $T_{ab(finite)}$ tends to zero for large $r$
in such a way as to remove divergences. It is this requirement for the
conserved quantities to be finite which uniquely determines the form of the
counterterms.

\section{Calculations}
\qquad Setting $S_{(ct)}=\int_{\partial 
B}\sqrt{|\gamma|}\mathcal{L}_{(ct)}$ the
following Lagrangian is required to remove divergences up to d=8
\cite{ref:krauslarsen}:
\begin{eqnarray}
\mathcal{L}_{(ct)}=-\alpha (d-1) - \frac{R}{2\alpha (d-2)} -
\frac{1}{\alpha^{3}2(d-2)^{2}(d-4)}\left(R^{ab}R_{ab}-\frac{d}{4(d-1)}R^2\right)
\nonumber \\ 
+\frac{1}{\alpha^{5}(d-2)^{3}(d-4)(d-6)} \left[
\frac{3d+2}{4(d-1)}RR^{ab}R_{ab}- \frac{d(d+2)}{16(d-1)^{2}}R^{3}-
2R^{ab}R^{cd}R_{acbd} \right.  \nonumber \\ 
\left.
+\frac{d-2}{2(d-1)}R^{ab}R_{;b;a}-R^{ab}R_{ab;c}^{\,\,\,\, ;c}+
\frac{1}{2(d-1)}R\,R_{\, ;c}^{;c}\right]. \label{eq:lagrangian}
\end{eqnarray}

The action is to be varied with respect to the boundary metric,
$\frac{\delta S_{(ct)}}{\delta\gamma_{ab}}$, and this yields the following
for the stress-energy counterterm:

\newpage
\begin{eqnarray}
T^{ab}_{(ct)}&=&-\alpha (d-1)\gamma^{ab} + \frac{1}{\alpha (d-2)}G^{ab}
+\frac{1}{\alpha^{3}(d-2)^{2}(d-4)
} \left[\frac{1}{2}\left(\frac{d}{4(d-1)}R^{2} -
R^{cd}R_{cd}\right)\gamma^{ab}\right. \nonumber \\
&-&\left.\frac{d}{2(d-1)} RR^{ab} + 
2R_{cd}R^{cadb}-\frac{d-2}{2(d-1)}R^{;a;b}+
R^{ab\,\,\,\, ;c}_{\,\,\, ;c} -\frac{1}{2(d-1)}R_{;c}^{\,\, 
;c}\gamma^{ab} \right] \nonumber \\
&+&\frac{2}{\alpha^{5}(d-2)^{3}(d-4)(d-6)}\left\{\frac{3d+2}{4(d-1)} 
\left[-G^{ab} R^{cd}R_{cd}\right.\right. \nonumber \\
&-&2RR^{ca}R^{b}_{\, c}+(R^{cd}R_{cd})^{;a;b}-\gamma^{ab} 
(R^{cd}R_{cd})_{;e}^{\, ;e} +2(RR^{bc})^{;a}_{\,\, ;c} - 
\gamma^{ab}(RR^{cd})_{;c;d} \nonumber \\
&-&\left. (RR^{ab})_{;c}^{\, ;c}\right] -\frac{d(d+2)}{16(d-1)^{2}} 
\left[\frac{1}{2}R^{3}\gamma^{ab}-3(R^{2}R^{ab}+(R^{2})^{;a;b}- 
(R^{2})^{;c}_{;c}\gamma^{ab}) \right] \nonumber \\
&-&2\left[\frac{1}{2}R^{ef}R^{cd}R_{ecfd}\gamma^{ab} -3 
R^{ae}R^{cd}R^{b}_{ced}+(R^{ac}R^{bd})_{;c;d} -(R^{ab}R^{cd})_{;c;d}
\right. \nonumber \\
&+& \left. 2(R^{b}_{cfd}R^{cd})^{;a;f} - 
(R_{fcgd}R^{cd})^{;g;f}\gamma^{ab}
-(R_{cd}R^{cadb})_{;e}^{;e} \right]\nonumber \\
&+&\frac{d-2}{2(d-1)}\left[\frac{1}{2}R^{cd}R_{;d;c}\gamma^{ab} 
-2R^{;(ca)}R^{b}_{c} +\frac{1}{2}\left[2R^{(;c;b);a}_{\;\;\;\;\;\;\;\;\; 
;c} + 
 -R^{;c;d}_{\,\,\,\;\;\, ;c;d}\gamma^{ab} \right.\right. \nonumber \\
&-&\left.\left.(R^{;b;a})_{;c}^{\, ;c}\right]+(R^{,a}R^{bc})_{;c}- 
\frac{1}{2}(R^{ab}R^{,c})_{;c} +(R^{cd}_{\,\,\, ;c;d})^{;a;b}\right] 
 \nonumber \\
&-&\left[\frac{1}{2}R^{cd}(R_{cd})_{;e}^{\, ;e} \gamma^{ab} 
+2(R^{bc\,e}_{\,\,\, ;e})^{;a}_{\, ,c} - (R^{ab\, ;c}_{\,\,\, 
;c})_{;e}^{\, ;e} -2R^{a\,\, ;c}_{\, ;c}R^{cb} -R^{cd}R_{cd}^{\,\,\, 
;a;b} \right. \nonumber \\
&-&\frac{1}{2}(R^{cd\, ;e}_{\,\,\, ;e})_{;c;d}\gamma^{ab} 
-(R^{ac;f}R^{b}_{c})_{;f} 
-2(R^{fc;a}R^{b}_{\, c})_{;f} + 2(R^{bc;a}R^{d}_{\, c})_{;d} + 
(R_{cd}^{\,\,\, ;a}R^{cd})^{;b} \nonumber \\
&-& \left. (R_{cd}^{\,\,\,;e} 
R^{cd})_{;e}\gamma^{ab} - 2 (R^{a}_{\, c})_{;e}^{\,\,\, ;e} R^{cb} + 
(R^{a}_{\,c}R^{bc;e})_{;e}\right] +\frac{1}{2(d-1)} \left[-G^{ab} 
R_{,c}^{\, 
;c} -R\, R^{,b;a} \right. \nonumber \\
&+& 2(R_{;c}^{\, ;c})^{;a;b} - 
2(R_{;c}^{\, ;c})_{;e}^{\, ;e} \gamma^{ab} 
+(R\,R^{,a})^{;b} -\frac{1}{2}(R\,R^{,c})_{;c} 
\gamma^{ab} +R_{,c}^{\, ;c}R^{ab} \nonumber \\
&+&\left.\left. (R^{cd}_{\,\,\, 
;c;d})_{;e}^{\,\, 
;e}\gamma^{ab}  \right]\right\} \label{eq:stresscounter}
\end{eqnarray}
with $G^{ab}$ being the Einstein tensor formed from the boundary metric 
$\gamma^{ab}$. This, albeit rather large, expression will allow us to 
compute the
conserved charges of the spacetime \footnote{For calculations of
conserved charges and Casimir energies of $d=4$ Kerr-AdS spacetimes see
\cite{ref:awad}}. We choose to study the higher
dimensional Schwarzschild-anti-deSitter black holes whose geometry in
Schwarzschild  coordinates is given by:
\begin{equation}
ds^{2}=-\left(1-\left(\frac{r_{0}}{r}\right)^{p-1}+\alpha^{2}r^{2}\right)dt^{2} +
\frac{dr^{2}}{\left(1-\left(\frac{r_{0}}{r}\right)^{p-1}+\alpha^{2}r^{2}\right)}+
r^{2}d\Omega^{2}_{p} \label{eq:schwads}
\end{equation}
where $p=d-1$. $d\Omega^2$ is the metric on unit $p$-spheres which, for
arbitrary $p$, is given by:
\begin{equation}
d\Omega_{p}^{2}= \left[d\theta_{0}^{2}+\sum_{n=1}^{p-1}d\theta_{n}^{2}
\left(\prod_{m=1}^{n}\sin^{2}\theta_{m-1}\right)\right]. \label{eq:psphere}
\end{equation}

\qquad Using (\ref{eq:energy}) we calculate the masses of 5, 7 and 9
dimensional black holes which may be summarized as follows:
\begin{eqnarray}
M_{5}&=&\frac{3\pi r_{0}^2}{8}+\frac{3\pi}{32\alpha^2} \nonumber \\
M_{7}&=&\frac{5\pi^{2} r_{0}^4}{16}-\frac{5\pi^{2}}{128\alpha^4} \nonumber \\
M_{9}&=&\frac{7\pi^{3} r_{0}^6}{48}+\frac{35\pi^{3}}{3072\alpha^6}. 
\label{eq:masses} \end{eqnarray}
In the limit of vanishing black hole mass, $r_{0}=0$, we have a pure vacuum
state which has non-zero energy. It was noted in \cite{ref:balakraus} and
\cite{ref:meyers2} that, in light of the AdS/CFT correspondence
\cite{ref:maldacena}, the second term in $M_{5}$ may be interpreted as 
the Casimir energy of the dual field theory which is $d=4$, 
$\mathcal{N}=4$, SUSY Yang Mills theory.
It is interesting to note that in seven dimensions the dual field theory 
has {\em negative} energy whereas the other cases yield positive energy. 
The dual field theory to $AdS_{7} \times S^{4}$ supergravity is the large 
N limit of the $d=6$ $(2,0)$ tensor multiplet theory \cite{ref:maldacena} 
of which little is known.
The field spectrum of this theory includes five scalars, a Majorana-Weyl 
spinor and a 2-form potential, $B$, with self-dual three form field 
strength, $dB$ and is the intrinsic theory on $N$ parallel M5 branes in 
the zero coupling limit. The calculation here seems to imply that the 
Casimir 
energy of such a theory on $S^{5}\times R$ is negative. We convert the 
above expression to gauge theoretic quantities via
\begin{equation}
\alpha^{-1}=2\mathit{l}_{p}(\pi N)^{1/3},
\end{equation}
with $\mathit{l}_{p}$ being the Planck length. This yields a Casimir mass of
\begin{equation}
E_{Casimir(2,0)}=-\frac{5}{8}\pi^{10/3}\mathit{l}_{p}^{4}N^{4/3}.
\end{equation}

\qquad Examples of negative energy solutions in general relativity are known 
such 
as the analytically continued Reissner-Nordstr\"{o}m solution (continued 
to both imaginary time and charge) although it is debatable how physical 
this construction is. Also, it has been noted that if one considers the 
Euclidean time 
extension of $D=d+1$ dimensional Schwarzschild-anti-deSitter black holes, 
the energy corresponding to the dual gauge theory on $S^{d-1}\times 
S^{1}$ is given by \cite{ref:horoandmeyers}:
\begin{equation}
E=-\frac{\Omega_{d-1}\beta r_{0}^{d-1}\alpha}{16\pi},
\end{equation}
where $\beta$ is the period of $S^{1}$ required to make the solution 
smooth at the (Euclidean) horizon. The negative sign arises from the 
supersymmetry breaking boundary conditions imposed along $S^{1}$. Since 
no such condition is imposed in the $D=7$ case where the boundary 
topology is $S^{5}\times R$ it is curious that a negative energy is 
produced. 

\subsection{Anomaly calculation}
\qquad Recent interesting work regarding lower dimensional anomalies 
may be found in \cite{ref:nojiri} and \cite{ref:ho} as well as 
\cite{ref:imbimbo} where diffeomorphism techniques are utilized. Taking 
the trace of the full stress-energy tensor yields: 
\begin{eqnarray}
T&=&\frac{1}{8\pi}\left[-(d-1)K-\alpha d(d-1)-\frac{R}{2\alpha} +
\frac{1}{\alpha^{3}2(d-2)^2}\left(\frac{d}{4(d-1)}R^{2}-R_{ab}R^{ab}
\right) \right. \nonumber \\
&+&\frac{1}{\alpha^{5}(d-2)^{3}(d-4)}\left(\frac{3d+2}{4(d-1)}RR^{ab}R_{ab} -
\frac{d(d+2)}{16(d-1)^2}R^{3}- 2R^{ab}R^{cd}R_{acbd} \right. \nonumber \\
&+&\left.\left.\frac{d-2}{2(d-1)}R^{ab}R_{;b;a}- R^{ab}R_{ab;c}^{;c} +
\frac{1}{2(d-1)}RR_{;c}^{;c}\right)\right]. \label{eq:trace}
\end{eqnarray}

\qquad We now wish to extract the Weyl anomaly from (\ref{eq:trace}). To
do this the metric must be expanded in a power series in $1/r$,
\begin{equation}
\gamma_{ab}=r^{2}\gamma_{(0)ab} + \gamma_{(2)ab} +
\frac{1}{r^{2}}\gamma_{(4)ab} + \ldots . \label{eq:expansion}
\end{equation}
(The Einstein equations dictate that no even power appears in the expansion),
The lowest order term in the
expression (\ref{eq:trace}) is then identified with the anomaly. This has
been shown to correspond with the work of Henningson and Skenderis
\cite{ref:hands} for $d \leq 6$. The $d=8$ case is very labour intensive
and will be addressed in a later revision. Instead, we utilize a
different method to calculate the anomaly which will also act as a check
for future calculations.

\qquad For the following, we adopt the coordinate system of
\cite{ref:hands}. This amounts to making the
transformation:
\begin{eqnarray}
r\rightarrow 1/\rho^{1/2} \nonumber \\
g_{ab}=\rho\gamma_{ab}. \label{eq:newexpansion}
\end{eqnarray}
Using
(\ref{eq:newexpansion}) the
effective, renormalized action may be obtained from the following density
\begin{equation}
\mathcal{L}=\alpha d \int_{\epsilon}d\rho \rho^{-d/2-1}\sqrt{g}
+\rho^{-d/2}\left(-2\alpha d\sqrt{g} +
4\alpha\rho\partial_{\rho}\sqrt{g}\right)|_{\rho=\epsilon}, \label{eq:newlag}
\end{equation}
where $\epsilon > 0$ is the cutoff point for the $\rho$ integration.

\qquad For even $d$ a logarithmic term appears from the integral which
arises
from the bulk part of the gravitational action, i.e. the usual Einstein -
Hilbert term with cosmological constant.
\begin{equation}
\mathcal{L}=\sqrt{g_{(0)}}\left[a_{(0)}\epsilon^{-d/2} +
a_{(2)}\epsilon^{-d/2+1} + \ldots + a_{(d-2)}\epsilon^{-1} - a_{(d)}\ln
(\epsilon)\right] + \hbox{finite terms}. \label{eq:serieslag}
\end{equation}
It is the coefficient of this logarithmic
term ($a_{(d)}$) which is to be identified with the anomaly. By expanding
$\sqrt{g}$
to order $\rho^{4}$ we may obtain the anomaly up to and including $d=8$.

\qquad The Einstein equations are given by \cite{ref:hands}
\begin{eqnarray}
\rho(2g^{\prime\prime} - 2g^{\prime}g^{-1}g^{\prime} + 
\tr[g^{-1}g^{\prime}]g^{\prime})
+\frac{1}{\alpha^{2}}\hbox{Ric}(g)-(d-2)g^{\prime} - \tr 
[g^{-1}g^{\prime}]g = 0 \nonumber \\
(g^{-1})^{ab}(g\prime_{ab;c} - g\prime_{ca;b})=0 \nonumber \\
\tr [g^{-1}g^{\prime\prime}]-\frac{1}{2}\tr [g^{-1}g^{\prime} g^{-1}
g^{\prime}]=0, \label{eq:einsteq}
\end{eqnarray}
where primes denote ordinary differentiation with respect to $\rho$ and
$\hbox{Tr}$ is the trace operator. All quantities are constructed with
respect to $g$.

\qquad  By using equations (\ref{eq:einsteq}) we can determine the
following relations between the $g$'s:
\begin{eqnarray}
\tr [g_{(0)}^{-1}g_{(4)}] &=&
\frac{1}{4}\tr\left[(g_{(0)}^{-1}g_{(2)})^{2}\right] \nonumber \\
\tr(g_{(0)}^{-1}g_{(6)}) &=&
\frac{2}{3}\tr\left[g_{(0)}^{-1}g_{(2)}g_{(0)}^{-1}g_{(4)}\right] -
\frac{1}{6}\tr\left[(g_{(0)}^{-1}g_{(2)})^{3}\right] \nonumber \\
\tr
[g_{(0)}^{-1}g_{(8)}]&=&\frac{1}{8}\tr\left[\left(
g_{(0)}^{-1}g_{(2)}\right)^{4}\right] +
\frac{3}{4}\tr\left[g_{(0)}^{-1}g_{(2)}g_{(0)}^{-1}g_{(6)}\right]
\nonumber \\
&-&
\frac{7}{12}\tr\left[\left(g_{(0)}^{-1}g_{(2)}\right)^{2}g_{(0)}^{-1}g_{(4)}
\right] + \frac{1}{3}\tr\left[\left(g_{(0)}^{-1}g_{4}\right)^{2}\right].
\label{eq:grels}
\end{eqnarray}
$g_{(2)ab}$ and $g_{(4)ab}$ may be found in \cite{ref:hands} and are
given here for reference.
\begin{eqnarray}
g_{(2)ab}&=&\frac{1}{\alpha^{2}(d-2)}\left(R_{(0)ab}-
\frac{1}{2(d-1)}R_{(0)}g_{(0)ab}\right) \nonumber \\
g_{(4)ab}&=&\frac{1}{\alpha^{4}(d-4)} \left(\frac{1}{4(d-2)}R_{(0)ab;c}^{;c}
-\frac{1}{8(d-1)}R_{(0);b;a} \right. \nonumber \\
&-&\frac{1}{8(d-1)(d-2)}R_{(0);c}^{;c}g_{(0)ab}-
\frac{1}{2(d-2)}R_{(0)}^{cd}R_{(0)acbd} \nonumber \\
&+&\frac{d-4}{2(d-2)^{2}}R^{c}_{(0)a}R_{(0)cb} +
\frac{1}{(d-1)(d-2)^{2}}R_{(0)}R_{(0)ab} \nonumber \\
&+& \left. \frac{1}{4(d-2)^{2}}R^{cd}_{(0)}R_{(0)cd}g_{(0)ab} -
\frac{3d}{16(d-1)^{2}(d-2)^{2}}R^{2}g_{(0)ab}, \right)
\end{eqnarray}
and we calculated $g_{6}$ to be:
\begin{eqnarray}
g_{(6)ab}&=&\frac{1}{3(6-d)}\left[4(g_{(2)}g_{(0)}^{-1}g_{(4)})_{ab} +
4(g_{(4)}g^{-1}_{(0)}g_{(2)})_{ab}- 2
\left(g_{(2)}g^{-1}_{(0)}\right)^{3}_{ab} \right. \nonumber \\
&-&\tr [g_{(0)}^{-1}g_{(2)}]g_{(4)ab} +
\tr\left[\ginv\gt\ginv\gf\right]g_{(0)ab}
+\frac{1}{2}\tr\left[\left(\ginv\gt\right)^{3}\right]g_{(0)ab} \nonumber \\
&-&\frac{1}{\alpha^{2}2}\left[\left[g_{(4)b;a}^{c}+g_{(4)a;b}^{c}-
g_{(4)ab}^{\,\,\, ;c} - g_{(2)}^{cd}\left(g_{(2)db;a} + g_{(2)ad;b} +
g_{(2)ab;d}\right)\right]_{;c}\right. \nonumber \\
&-&\left[\left(\tr[\ginv\gf ]\right)_{;b} -
\gt^{cd}g_{(2)cd;b}\right]_{;a} +\frac{1}{2}\left[\left[\tr
[\ginv\gt]
\left(g_{(2)b;a}^{c} + g_{(2)a;b}^{c} + g_{(2)ab}^{;c}\right)\right]
_{;c} \right. \nonumber \\
&-&\left.\left. g^{c}_{(2)d;a} g^{d}_{(2)c;b} - 
g_{(2)ca;d}g^{d\,\,\,\,\,\,\, ;c}_{(2)b} +2
g_{(2)ca}^{\,\,\,\,\,\,\,\, ;d}g_{(2)b;d}^{c}\right]\right]. \label{eq:gsix}
\end{eqnarray}

\qquad We now use the expansion of $g$ to order $\rho^{4}$ to calculate
the anomaly for $d=8$. To this order:
\begin{eqnarray}
\sqrt{g}=\sqrt{\gz}\left[ 1+\frac{1}{2}\tr A -\frac{1}{4} \tr A^{2}
+\frac{1}{6}\tr A^{3} - \frac{1}{8}\tr A^{4} + \frac{1}{8} (\tr A)^{2} \right.
\nonumber \\ -\frac{1}{8}\tr A\tr A^{2} + \frac{1}{12}\tr A\tr A^{3}
+\frac{1}{32} (\tr A^{2})^{2} \nonumber \\ \left. +\frac{1}{48}(\tr A)^3
-\frac{1}{32}(\tr A)^{2}\tr A^{2} + \frac{1}{384}(\tr A)^{4} \right],
\end{eqnarray}
where
\begin{equation}
A\equiv \rho\ginv\gt + \rho^{2}\ginv\gf +\rho^{3}\ginv\gs +\rho^{4}\ginv
g_{(8)}.
\end{equation}
The anomaly is now given by studying terms of
$\mathcal{O}\left(\rho^{4}\right)$ and using (\ref{eq:grels}):
\begin{eqnarray}
a_{(8)}&=& -\frac{1}{16}\tr[(\ginv\gt )^{4}] -
\frac{1}{8}\tr[\ginv\gt\ginv\gs ] + \frac{5}{24}\tr[(\ginv\gt
)^{2}\ginv\gf ] \nonumber \\
&-&\frac{1}{12}\tr[(\ginv\gf)^{2}]
-\frac{1}{12}\tr[\ginv\gt]\tr[\ginv\gt\ginv\gf] \nonumber \\
&+&\frac{1}{24}\tr[\ginv\gt]\tr[(\ginv\gt)^{3}]
+\frac{1}{128}(\tr[(\ginv\gt)^{2}])^{2} \nonumber \\
&-&\frac{1}{64}(\tr[\ginv\gt])^{2}\tr[(\ginv\gt)^{2}] 
+ \frac{1}{384}(\tr[\ginv\gt])^{4}. \label{eq:anom}
\end{eqnarray}

\qquad In general, up to constant coefficients, the anomaly to $d$ 
dimension is given by all combinations of $\tr A$ up to $A^{d/2}$. 

\section{Conclusion}

\qquad We considered here a counterterm subtraction technique to study 
stress-energy tensors in higher dimensional gravity. We find that the 
counterterm method consistently produces correct black hole masses for $d 
\leq 8$ and therefore this is a most useful technique even when there is no 
reference background with which to renormalize the energy. The 
calculations also produce non-zero vacuum energies in the $r_{0}=0$ 
limit. These are interpreted as Casimir energies of the boundary field 
theory \cite{ref:balakraus} which includes a {\em negative} energy 
contribution. We also find, by identifying logarithmic 
divergences in the gravitational action, the Weyl anomaly in the $d=8$ 
case. It will be interesting to see if the counterterm technique produces 
a similar expression for the anomaly.

\section{Acknowledgements}
The authors would like to thank Dr. P. Kraus and Dr. F. Larsen of the
University of Chicago for helpful clarifying comments via e-mail. Some 
calculations were performed using the MAPLE$^{\hbox{\small TM}}$ tensor 
package.

\newpage
\bibliographystyle{unsrt}

\end{document}